\documentclass[reprint,preprintnumbers,onecolumn]{revtex4-1} 
\usepackage{amsmath,bm}  % needed for \tfrac, \bmatrix, etc.
\usepackage{amsfonts} % needed for bold Greek, Fraktur, and blackboard bold
\usepackage{graphicx} % needed for figures
\usepackage{mathrsfs} 
\usepackage{amssymb}
\usepackage{dsfont}
\usepackage{bbm}
\usepackage[cal=boondox,scr=boondoxo]{mathalfa}
\usepackage[bbgreekl]{mathbbol}
\usepackage{hyperref}
\usepackage{relsize}% enlargement
\usepackage{color}
\usepackage{bbold}
\begin{document}
\preprint{\textsl{KPOP}$\mathscr{E}$-2020-07}
% Be sure to use the \title, \author, \affiliation, and \abstract macros
% to format your title page.  Don't use lower-level macros to  manually
% adjust the fonts and centering.

\title{Derivation of Jacobian Formula
with Dirac Delta Function
}
% In a long title you can use \\ to force a line break at a certain location.
\author{Dohyun Kim}
\email{tradysori705@korea.ac.kr} 
\author{June-Haak Ee}
\email{chodigi@gmail.com} 
\author{Chaehyun Yu}
\email{chyu@korea.ac.kr} 
\author{Jungil Lee}
\email{jungil@korea.ac.kr} % optional
\thanks{Director of the Korea Pragmatist Organization for Physics Education (\textsl{KPOP}$\mathscr{E}$)}
\affiliation{\textsl{KPOP}$\mathscr{E}$ Collaboration, Department of Physics, Korea University, Seoul 02841, Korea}
%\altaffiliation[permanent address: ]{101 Main Street, 
%  Anytown, USA} % optional second address
% If there were a second author at the same address, we would put another 
% \author{} statement here.  Don't combine multiple authors in a single
% \author statement.
% Please provide a full mailing address here.

% See the REVTeX documentation for more examples of author and affiliation lists.
%\date{\today}

\begin{abstract}
We demonstrate how to make
 the coordinate transformation or change of variables
from Cartesian coordinates to curvilinear coordinates by making use of a convolution
of a function with Dirac delta functions whose arguments are determined by the transformation
functions between the two coordinate systems.
By integrating out an original coordinate with a Dirac delta function, we replace the original coordinate with a new coordinate in a systematic way.
A recursive use of Dirac delta functions allows the coordinate
transformation successively. 
After replacing 
every original coordinate into a new curvilinear coordinate,
we find that the resultant Jacobian of the corresponding coordinate transformation is automatically obtained in a completely algebraic way.
In order to provide insights on this method, we present a few examples
of evaluating the Jacobian explicitly without resort to 
the known general formula.

\end{abstract}
% AJP requires an abstract for all regular article submissions.
% Abstracts are optional for submissions to the "Notes and Discussions" section.

\maketitle % title page is now complete

%\begin{widetext}
\section{Introduction} % Section titles are automatically converted to all-caps.
% Section numbering is automatic.
A coordinate transformation or change of variables from a coordinate system to another in multi-dimensional integrals 
has widely been applied to a variety of fields in mathematics and physics. 
This transformation always involves a factor called the \textit{Jacobian}, which is the determinant of 
the Jacobian matrix. The matrix elements of the Jacobian matrix are the first-order partial derivatives
of the new coordinates with respect to the original coordinates.
The formula for the change of variables from $n$-dimensional variables $x_1$, $x_2$, $\cdots$, $x_n$ to $q_1$, $q_2$, 
$\cdots$, $q_n$  is expressed in terms of the Jacobian $\mathscr{J}$:
\begin{equation}
\int d x_1 d x_2 \cdots d x_n \,f(x_1,\cdots,x_n) = \int d q_1 d q_2 \cdots d q_n \,\mathscr{J} \,F(q_1,\cdots,q_n),
\end{equation}
where the integrand $f(x_1,\cdots,x_n)$ is a function of 
 the independent variables  $x_1$, $\cdots$, $x_n$
 and $F(q_1,\cdots,q_n)=f[x_1(q_1,\cdots,q_n),\cdots, x_n(q_1,\cdots,q_n)]$.
 In general, the variables $x_1$, $\cdots$, $x_n$ can be treated as the  
 Cartesian coordinates of an $n$-dimensional Euclidean space
 and the variables $q_1$, $q_2$, $\cdots$, $q_n$ form a set of
 curvilinear coordinates representing the same Euclidean space.
 We assume that each curvilinear coordinate $q_i$ is uniquely defined by
 the Cartesian coordinates: $q_i=q_i(x_1,\cdots,x_n)$. We also assume that
 the inverse transformation
 is uniquely defined as $x_i=x_i(q_1,\cdots,q_n).$
Then the Jacobian $\mathscr{J}$ can be expressed as
\begin{equation}
\mathscr{J} = 
\left|
\frac{\partial (x_1, x_2, \cdots, x_n)}{\partial(q_1, q_2, \cdots, q_n)}
\right|
=
\left|
\mathscr{Det}
\begin{pmatrix}
\displaystyle
\frac{\partial x_1}{\partial q_1}&
\displaystyle\frac{\partial x_2}{\partial q_1}&
\cdots &
\displaystyle\frac{\partial x_n}{\partial q_1}
\\
\displaystyle\frac{\partial x_1}{\partial q_2}&
\displaystyle\frac{\partial x_2}{\partial q_2}&
\cdots &
\displaystyle\frac{\partial x_n}{\partial q_2}
\\
\vdots & \vdots & \ddots & \vdots 
\\
\displaystyle\frac{\partial x_1}{\partial q_n}&
\displaystyle\frac{\partial x_2}{\partial q_n}&
\cdots &
\displaystyle\frac{\partial x_n}{\partial q_n}
\end{pmatrix}
\right|,
\label{jacobian}%
\end{equation} 
where $\mathscr{Det}$ stands for the determinant.

In physics, the Jacobian appears frequently when one makes the change of variables 
between Cartesian and curvilinear coordinates in various physical quantities involving surface or volume integrals.
However, in most physics textbook including classical mechanics and electromagnetism 
usually abstract descriptions are provided. 
In many textbooks of calculus or mathematical physics~\cite{arfken},
the Jacobian formula is derived in the following way:
First, one transforms the multivariable differential volume by applying the change of variables.
Next, one imposes a geometrical argument that the infinitesimal volume
is invariant under the transformation~\cite{stewart}.
The invariance of the volume can also be confirmed by applying Green's theorem to show that
$\int_S d x_1 d x_2 = \int_T \mathscr{J} d q_1 d q_2,$ where $S$ is a rectangular region 
in the $x_1 x_2$ plane and $T$ is the corresponding region~\cite{apostol}.
Although experienced teachers or researchers may follow the abstract logic in this kind of Jacobian derivation
without difficulties, the concept is rather unclear or less intuitive to undergraduate physics-major students 
who are not familiar with advanced mathematical concepts of multi-variable calculus.

The Dirac delta function $\delta(x)$ is not a well-defined function but a distribution defined
only through integration:
$\int_{-\infty}^{\infty} d x \, \delta (x) = 1$.
For any smooth function $f(x)$,
$\int_{-\infty}^{\infty} d x \,\delta (x - y) f(x) = f(y)$.
This property can be applied to changing the integration variable.
Recently, some of us introduced an alternative proof of
Cramer's rule by making use of Dirac delta functions~\cite{cramer}.
It turns out that the method with a convolution of a coordinate vector with  Dirac delta functions provides a systematic way to change integration variables from original coordinates to new coordinates.
This change of variables enables us to reproduce Cramer's rule.

The Dirac delta function technique exploited in the derivation of Cramer's rule 
in Ref.~\cite{cramer} can be immediately applicable to the evaluation of
the Jacobian for the coordinate transformation or change of variables
after replacing the coordinate vector with an arbitrary function.
However, the direct application of the approach in Ref.~\cite{cramer} 
is limited to a linear transformation  between two coordinate systems because Cramer's rule applies
only to the system of linear equations. Thus the method is not applicable to the transformation
involving a set of curvilinear coordinates which are frequently used in physics.

The main goal of this work is to present a more intuitive derivation of the Jacobian 
involving any coordinate transformation and to demonstrate how it works with heuristic examples. 
We develop an alternative derivation of the Jacobian formula as well as the coordinate
transformation by convolving a function with Dirac delta functions.
Our derivation relies only on the direct integration of Dirac delta functions whose arguments involve the coordinate transformation rules.
Hence, we expect that students who are familiar with the Dirac delta function can compute
the Jacobian formula in any coordinate transformations by themselves without referring to a reference. 
The approach that we present in this paper is quite straightforward and requires mostly algebraic computation
skills.

In this paper, we derive the Jacobian for a coordinate transformation or
change of variables in the case of a non-linear transformation
by convolving an arbitrary function with Dirac delta functions.
While the basic strategy to perform the coordinate transformation is similar to
that employed in Ref.~\cite{cramer}, 
 the integration of the original coordinates is rather involved
  because of the non-linear property of the transformation functions between the two coordinate systems.
The integration of the original coordinates can be carried out by making use of
Dirac delta functions. An extra factor containing partial derivatives of corresponding coordinate 
variables appears in front of the original integrand after that multiple integration.
It turns out that the extra factor  can be evaluated
by making use of the chain rule of partial derivatives.
A recursive use of Dirac delta functions enables us to achieve the coordinate 
transformation successively.
After replacing every coordinate with new coordinates, we identify 
the resultant extra factor in the integrand with the Jacobian for the coordinate
transformation or change of variables.

The derivation of the Jacobian formula presented in this paper is new to our best knowledge.
It is remarkable that our derivation is free of borrowing
abstract and advanced mathematical concepts unlike the other derivations available.
Instead, we exploit a simple concept of integration of the one-dimensional Dirac delta function 
repeatedly in combination with a purely
algebraic manipulation in reorganizing the extra factor by applying chain rules.
This intuitive and systematic approach is expected to be pedagogically useful 
in upper-level mathematics or physics courses
in practice of  the recursive use of both the Dirac delta function
and the chain rule of partial derivatives.

This paper is organized as follows:
In section~\ref{sec;derivation},
we introduce notations that are frequently used in the rest of the paper,
make a rough sketch of our strategy and
present a formal derivation of the Jacobian factor for the coordinate transformation from the 
$n$-dimensional Cartesian coordinates to a set of curvilinear coordinates.
Section~\ref{sec;app} is devoted to the explicit evaluation
of the Jacobians for a few examples.
Conclusions are given in section~\ref{sec;con} and
a rigorous derivation of
the chain rule for partial derivatives is given in Appendices.

%%%%%%%%%%%%%%%%%%%%%%%%%%%%%%%%%%%%%%%%%%%%%%%%%%%%%%%%%%%%%
\section{Derivation of the Jacobian  \label{sec;derivation}}
%%%%%%%%%%%%%%%%%%%%%%%%%%%%%%%%%%%%%%%%%%%%%%%%%%%%%%%%%%%%%

\subsection{Strategy and Notation\label{subsec:SN}}

In this subsection, we present our strategy to derive
the Jacobian for a coordinate transformation or change of variables from the Cartesian coordinates
$x_i$
to the curvilinear coordinates $q_i$
with transformation functions
\begin{equation}
\label{qi}
q_i=q_i(x_1,\cdots,x_n)
\end{equation}
 for $i=1$ through $n$, where $n$ is a positive integer.
We assume that the two sets of coordinates describe a single point uniquely and, therefore,
the two sets of coordinates have a one-to-one correspondence although the transformation is
in general non-linear. Thus the transformation in  Eq.~\eqref{qi}
is invertible: the inverse transformation
from the curvilinear coordinates to the Cartesian coordinates exists.
If the curvilinear coordinates are a linear combination of the Cartesian coordinates,
then the linear transformation is invertible if the transformation matrix is
non-singular: the determinant of the matrix is not vanishing.  
Then, the inverse transformation can be written as
\begin{equation}
x_i=x_i(q_1,\cdots,q_n).
\end{equation}

The basic strategy to derive the Jacobian with Dirac delta functions is the same as that 
for the derivation of Cramer's rule for a partial set of a coordinate transformation or 
change of variables given in Ref.~\cite{cramer}. 
One could immediately apply the approach in Ref.~\cite{cramer} to find the Jacobian as long as
the transformation \eqref{qi} is linear.
In general, the transformation functions \eqref{qi} are non-linear. Here we develop a generalized
version of the approach in Ref.~\cite{cramer} in order to consider  
arbitrary curvilinear coordinates.

We define an $n$-dimensional integral $I_n$,
\begin{equation}
\label{In}
I_n=\int_{-\infty}^\infty\,d^n x\,f(x_1,\cdots,x_n),
\end{equation}
where $d^n x\equiv dx_1\cdots dx_n$ is the $n$-dimensional differential volume element
and
the integrand $f(x_1,\cdots,x_n)$ is an arbitrary function of the Cartesian coordinates.
We assume that every Cartesian coordinate is integrated over the region $(-\infty,\infty)$. 
We define the unity
\begin{equation}
\mathbbm{1}_i\equiv\int\,dq_i\,\delta[q_i-q_i(x_1,\cdots,x_n)]=1
\label{trivial}
\end{equation}
for $i=1$ through $n$.
Multiplying the unity $\mathbbm{1}_i$ to the integral $I_n$,
one can integrate out the integration variable $x_i$ by making use of the Dirac delta function
keeping the $q_i$ integral unevaluated.
By applying this process to $I_n$ recursively
from $i=1$ through $n$, we complete the change of the integration variables 
from the Cartesian coordinates to the curvilinear coordinates.
While we have suppressed the bounds of the integration for the curvilinear coordinate $q_i$'s
in Eq.~\eqref{trivial}, the curvilinear coordinates are assumed to be integrated over
the  entire region to cover the whole Euclidean space represented by the Cartesian coordinates
by a single time.

We first compute $\mathbbm{1}_1 \times I_n$:
\begin{equation}
\label{Ind}
I_n=\mathbbm{1}_1\times I_n=
\int\,d q_1
\int_{-\infty}^\infty\,d^n x\,
f(x_1,\cdots,x_n)
\delta[q_1-q_1(x_1,\cdots,x_n)].
\end{equation}
By definition, we integrate over $x_1$ by making use of
the delta function $\delta[q_1-q_1(x_1,\cdots,x_n)]$ keeping the $q_1$ integral unevaluated:
\begin{equation}
I_n=
\int\,d q_1
\int_{-\infty}^\infty\,d x_2 \cdots d x_n\,
\frac{1}{\mathscr{G}_1} f(q_1,x_2,\cdots,x_n),
\label{Inx1}
\end{equation}
where
$x_1$ is expressed in terms of
$q_1$ and $x_j$'s for $j=2$ through $n$
satisfying the condition that the argument of the delta function vanishes, $q_1-q_1(x_1,\cdots,x_n)=0$.
Because the explicit forms of $f(x_i)$ and $x_i$ vary depending on the integration step,
we adopt the notation $f(q_1,x_2,\cdots,x_n)$
after the $x_1$ integration.
We will present more detailed explanations for this notation in the later part of this subsection.
The extra factor $\mathscr{G}_1$ is the remnant of
the integration of the delta function and 
its explicit form will be given later in this paper.

Next, we multiply $\mathbbm{1}_2$ to $I_n$ in Eq. \eqref{Inx1} to find that
\begin{equation}
I_n=\mathbbm{1}_2\times I_n=\int\,d q_1 \,dq_2
\int_{-\infty}^\infty\,d x_2 \cdots d x_n\,
\frac{1}{\mathscr{G}_1}
f(q_1,\cdots,x_n)
\delta[q_2-q_2(q_1,x_2,\cdots,x_n)],
\end{equation}
where every $x_1$ in the argument of the delta function
as well as the integrand function is replaced 
with the expression in terms of 
$q_1$ and $x_j$'s for $j=2$ through $n$.
Performing the integration over $x_2$ by making use
of the delta function, we find
\begin{equation}
I_n=
\int\,d q_1 \,dq_2
\int_{-\infty}^\infty\,d x_3 \cdots d x_n\,
\frac{1}{\mathscr{G}_2}
f(q_1,q_2,x_3,\cdots,x_n).
\end{equation}
After the $x_2$ integration every
$x_2$ in the integrand is expressed in terms of $q_1$,
$q_2$ and $x_j$'s for $j=3$ through $n$.
Again, $\mathscr{G}_2$ is the remnant of the integration of the delta functions.

In this way, we integrate over $x_k$ for
$k=1$ through $n$ successively.
Finally, after the integration over $x_n$ we find that $I_n$ reduces into the $n$-dimensional multiple integral
over the curvilinear coordinates $q_i$  
for $i=1$ through $n$ only. 
\begin{equation}
I_n=
\int\,d q_1 \cdots dq_n\,
\frac{1}{\mathscr{G}_n}
f(q_1,\cdots,q_n),
\end{equation}
where $\mathscr{G}_n$ is the remnant of the integration
of the delta functions.
At this stage, the integrand
acquires an additional factor $\mathscr{G}_n$ in front of the original integrand $f$ in Eq.~\eqref{In}.
This extra factor is identified with the Jacobian.

In an intermediate step, for example, 
where the integration over $x_j$ ($1\le j \le n$) is carried out, 
$x_1, \cdots, x_j$ in the integrand 
must be replaced with the expressions in terms of
$q_1, \cdots, q_j$ and $x_{j+1}, \cdots, x_n$ as
\begin{eqnarray}
\label{xk}
x_k&=&x_k(q_1,\cdots,q_j,x_{j+1},\cdots,x_n)
\end{eqnarray}
for $k=1$ through $j$.
Each $x_k$ in Eq.~\eqref{xk} is determined by the condition
that the argument of the corresponding Dirac delta function vanishes:
\begin{eqnarray}
q_k-q_k[x_1(q_1,\cdots,q_j,x_{j+1},\cdots,x_n),\cdots,x_j(q_1,\cdots,q_j,x_{j+1},\cdots,x_n),x_{j+1},\cdots,x_n]&=&0.
\end{eqnarray}
One must keep in mind that
the explicit form of each $x_k$ varies depending on the integration step
as is displayed in Eq.~\eqref{xk}.
Thus, one must distinguish, for example,
$x_k(q_1,\cdots,q_j,x_{j+1},\cdots,x_n)$
from $x_k(q_1,\cdots,q_{j-1},x_j,\cdots,x_n)$,
where the former is the expression after the integration over
$x_j$ and the latter is that after the
integration over $x_{j-1}$.
This notation is also applied to the original integrand function
$f$ and the extra factor $\mathscr{G}_i$.

As an explicit example, we consider
the coordinate transformation between the 2-dimensional
Cartesian coordinates and the polar coordinates with
the transformation functions
\begin{equation}
r=\sqrt{x^2+y^2}\quad\textrm{and}\quad\theta=\arctan\dfrac{y}{x}
\end{equation}
and the inverse transformation functions
\begin{equation}
x=r\cos\theta\quad\textrm{and}\quad y=r\sin\theta.
\end{equation}
In an intermediate step, $y$ can be expressed in terms of $\theta$ and $x$ as
$y(\theta,x)=x\tan\theta$, which must be distinguished from
$y(\theta,r)=r\sin\theta$.

Since the dependence of a coordinate variable varies  according 
to the integration step, one must take special care in dealing with their partial derivatives.
In order to avoid such an ambiguity, we introduce a notation for the partial derivative
with subscripts of variables that are held constant.
In the above $2$-dimensional transformation,
the partial derivative of $\theta$ with respect to $y$ holding $x$ fixed is denoted by
\begin{equation}
\label{D-theta-x}
\left(\frac{\partial \theta}{\partial y}\right)_{x}=
\frac{x}{x^2+y^2},
\end{equation}
while the partial derivative of $\theta$ with respect to $y$ holding $r$ fixed is represented by
\begin{equation}
\label{D-theta-r}
\left(\frac{\partial \theta}{\partial y}\right)_{r}=
\frac{1}{\sqrt{r^2-y^2}}.
\end{equation}
It is apparent from Eqs.~\eqref{D-theta-x} and \eqref{D-theta-r} that 
\begin{equation}
\left(\frac{\partial \theta}{\partial y}\right)_{x}
\neq
\left(\frac{\partial \theta}{\partial y}\right)_{r}.
\end{equation}

In a general case, for a variable $q_j (q_1,\cdots,q_i,x_{i+1},\cdots,x_n)$,
 we denote the partial derivative of $q_j$
with respect to $x_a$ holding $q_1, \cdots, q_i, x_{i+1}, \cdots, x_n$ fixed
 with the subscript $(i)$ as
\begin{equation}
\label{eq:pd-qx}
\left(\frac{\partial q_j}{\partial x_{a}}\right)_{(i)}=\frac{\partial q_j(q_1,\cdots,q_i,x_{i+1},\cdots,x_n)}{\partial x_{a}},
\end{equation}
where $i+1\le j\le n$ and $i+1\le a \le n$.
Similarly, for a variable $x_k(q_1,\cdots,q_i,x_{i+1},\cdots,x_n)$,
 the partial derivative of $x_k$ with respect to $q_{b}$ holding $q_1$, $\cdots$, $q_i$, $x_{i+1}$, $\cdots$, $x_n$ fixed  is denoted by
\begin{equation}
\label{eq:pd-xq}
\left(\frac{\partial x_k}{\partial q_{b}}\right)_{(i)}=\frac{\partial x_k(q_1,\cdots,q_i,x_{i+1},\cdots,x_n)}{\partial q_{b}},
\end{equation}
where $1\leq k\leq i$ and $1\leq b\leq i$.
We denote the partial derivative of $x_{\ell}$ with respect to $x_c$ holding $q_1$, $\cdots$, $q_i$, $x_{i+1}$, $\cdots$, $x_n$ fixed by
\begin{equation}
\label{eq:pd-xx}
\left(\frac{\partial x_{\ell}}{\partial x_c}\right)_{(i)}=\frac{\partial x_{\ell}(q_1,\cdots,q_i,x_{i+1},\cdots,x_n)}{\partial x_c},
\end{equation}
where $1\leq\ell\leq i$ and $i+1\leq c\leq n$.
Finally, we express the partial derivative of $q_{\ell}$ with respect to $x_c$ holding $x_1$, $\cdots$, $x_n$ without subscript:
\begin{equation}
\label{eq:pd-qxx}
\frac{\partial q_{\ell}}{\partial x_c}=\frac{\partial q_{\ell}(x_1, \cdots, x_n)}{\partial x_c},
\end{equation}
where $1\leq  {\ell}\leq n$ and $1\leq  c\leq n$.

In the coordinate transformation from $(x_1,\cdots,x_n)$
to $(q_1,\cdots,q_n)$,
we define the function $\mathscr{G}_k$, which is relevant for a partial set of integral variables 
corresponding to a transformation from
$(x_1,\cdots,x_k$) to ($q_1,\cdots,q_k$),  as
\begin{equation}
\mathscr{G}_k = 
\left|
\mathscr{Det}
\begin{pmatrix}

\dfrac{\partial q_1}{\partial x_1}

&

\dfrac{\partial q_2}{\partial x_1}

&
\cdots &

\dfrac{\partial q_k}{\partial x_1}

\\

\dfrac{\partial q_1}{\partial x_2}

&

\dfrac{\partial q_2}{\partial x_2}

&
\cdots &

\dfrac{\partial q_k}{\partial x_2}

\\
\vdots & \vdots & \ddots & \vdots 
\\

\dfrac{\partial q_1}{\partial x_k}

&

\dfrac{\partial q_2}{\partial x_k}

&
\cdots 
&

\dfrac{\partial q_k}{\partial x_k}

\end{pmatrix}
\right|
\end{equation} 
for $k=1$ through $n$.
Note that $\mathscr{G}_n$ is the inverse of the Jacobian $\mathscr{J}_n$ which is defined by
\begin{equation}
\mathscr{J}_n = 
\left|
\mathscr{Det}
\begin{pmatrix}
\left(\dfrac{\partial x_1}{\partial q_1}\right)_{(n)}&
\left(\dfrac{\partial x_2}{\partial q_1}\right)_{(n)}&
\cdots &
\left(\dfrac{\partial x_k}{\partial q_1}\right)_{(n)}
\\
\left(\dfrac{\partial x_1}{\partial q_2}\right)_{(n)}&
\left(\dfrac{\partial x_2}{\partial q_2}\right)_{(n)}&
\cdots &
\left(\dfrac{\partial x_k}{\partial q_2}\right)_{(n)}
\\
\vdots & \vdots & \ddots & \vdots 
\\
\left(\dfrac{\partial x_1}{\partial q_k}\right)_{(n)}&
\left(\dfrac{\partial x_2}{\partial q_k}\right)_{(n)}&
\cdots &
\left(\dfrac{\partial x_k}{\partial q_k}\right)_{(n)}
\end{pmatrix}
\right|.
\end{equation}
 
%============================================================
\subsection{$\textbf{2}$-dimensional case}
%============================================================
In this subsection, we consider a $2$-dimensional integral $I_2$ for the integration
of an arbitrary function $f(x_1,x_2)$:
\begin{equation}
I_2=\int_{-\infty}^\infty\,dx_1 dx_2\,f(x_1,x_2).
\end{equation}
The Cartesian coordinates $x_1$ and $x_2$ are transformed
into curvilinear coordinates $q_1$ and $q_2$ with
the transformation relations
\begin{equation}
q_1=q_1(x_1,x_2), \quad q_2=q_2(x_1,x_2),
\end{equation}
which are assumed to be invertible and non-singular.
Thus, $x_1$ and $x_2$ can be expressed in terms of $q_1$ and $q_2$:
\begin{equation}
x_1=x_1(q_1,q_2), \quad x_2=x_2(q_1,q_2).
\end{equation}
We multiply the unities
\begin{equation}
\label{deltaqi}
\mathbbm{1}_i=\int dq_i\,\delta[q_i-q_i(x_1,x_2)]=1
\end{equation}
to $I_2$ for $i=1$,  $2$, sequentially.
First, we multiply $\mathbbm{1}_1$ in Eq.~\eqref{deltaqi} to $I_2$. Then, $I_2$ can be expressed as
\begin{equation}
\label{I2d2}
I_2 = \int \, d q_1\, \int_{-\infty}^\infty\,dx_1 dx_2\,
\delta[q_1-q_1(x_1,x_2)]
f(x_1,x_2).
\end{equation}

The integration over $x_1$ can be carried out by making use of
the delta function for $q_1$ as
\begin{eqnarray}
\label{x1d2}
\int_{-\infty}^\infty\,dx_1\,\delta[q_1-q_1(x_1,x_2)]
&=&
\frac{1}{\left|
\dfrac{\partial q_1}{\partial x_1}
\right|_{x_1=x_1(q_1,x_2)}}
\nonumber\\
&=&\frac{1}{\mathscr{G}_1(q_1, x_2)},
\end{eqnarray}
which leads to
\begin{equation}
\label{I2afterx1}
I_2 =\int \, d q_1\, \int_{-\infty}^\infty\,dx_2\,
\frac{1}{\mathscr{G}_1(q_1, x_2)}
f(q_1,x_2).
\end{equation}
After the integration over $x_1$, every
$x_1$ on the right-hand side of Eq.~\eqref{x1d2} and $\delta[q_2-q_2(x_1,x_2)] f(x_1,x_2)$
in Eq.~\eqref{deltaqi} must be replaced with $x_1(q_1,x_2)$ satisfying
the condition that the argument of the delta function vanishes, $q_1-q_1(x_1,x_2)=0$.
Then the delta function for $q_2$ and $f(x_1,x_2)$ can be expressed as
$\delta[q_2 - q_2(q_1,x_2)]$
and $f(q_1,x_2)$, respectively.

After multiplying $\mathbbm{1}_2$ in Eq.~\eqref{deltaqi} to $I_2$ in Eq.~\eqref{I2afterx1}, the integration over $x_2$ can be performed by making use of the remaining delta function for $q_2$ as
\begin{eqnarray}
\label{x2d2}
\int_{-\infty}^\infty\,dx_2\,\delta[q_2-q_2(q_1,x_2)]&=&\frac{1}{\left|\left(\dfrac{\partial q_2}{\partial x_2}\right)_{(1)}\right|_{x_2=x_2(q_1,q_2)}}
\nonumber\\
&=&\frac{\mathscr{G}_1(q_1, q_2)}{\mathscr{G}_2(q_1, q_2)}.
\end{eqnarray}
After the integration over $x_2$, every $x_2$ on the right-hand sides of Eqs.~\eqref{x1d2} and \eqref{x2d2} and in $f(q_1,x_2)$ is replaced with $x_2(q_1,q_2)$,
which can be obtained from the condition that the argument of the delta function vanishes,
$q_2-q_2(q_1,x_2)=0$. Then, we can express $x_1$ in terms of $q_1$ and $q_2$ 
by replacing $x_2$ with $x_2(q_1,q_2)$ in $x_1(q_1,x_2)$.
We have obtained the last equality of Eq.~\eqref{x2d2} by making use of the chain rule
for the partial derivatives. A rigorous proof of this formula is given in Eq.~\eqref{j2j1} 
of Appendix \ref{APA}.

After both $x_1$ and $x_2$ are integrated out, the integral $I_2$ reduces into
\begin{eqnarray}
\label{I2int}
I_2 
&=&
\int\,dq_1\,dq_2\,
\frac{1}{\mathscr{G}_1}
\times
\frac{\mathscr{G}_1}{\mathscr{G}_2} 
f[x_1(q_1,q_2),x_2(q_1,q_2)] 
=
\int\,dq_1\,dq_2\,
\mathscr{J}_2\,
f[x_1(q_1,q_2),x_2(q_1,q_2)],
\end{eqnarray}
where the Jacobian for the change of variables
is identified as $\mathscr{J}=\mathscr{J}_2=1/\mathscr{G}_2$.
This completes the proof of the Jacobian for a $2$-dimensional coordinate transformation or
change of variables.

\subsection{$\textbf{3}$-dimensional case \label{3dcase}}

In this subsection, we extend the results in the previous subsection
to the $3$-dimensional case.
This is a special case of the $n$-dimensional coordinate transformation or change of variables,
which we will prove in the next subsection.
However, it is worthwhile to prove the $3$-dimensional case in detail
for a pedagogical purpose.

We consider a $3$-dimensional integral $I_3$ for an arbitrary function
$f(x_1,x_2,x_3)$:
\begin{equation}
I_3=
\int_{-\infty}^\infty
\,dx_1
\,dx_2
\,dx_3
\,f(x_1,x_2,x_3).
\end{equation}
The Cartesian coordinates $x_i$ for $i=1$ through $3$ are transformed
into the curvilinear coordinates $q_i$'s as
\begin{equation}
q_1=q_1(x_1,x_2,x_3),\quad q_2=q_2(x_1,x_2,x_3),\quad q_3=q_3(x_1,x_2,x_3).
\end{equation}
The inverse transformation can be expressed as
\begin{equation}
x_1=x_1(q_1,q_2,q_3),\quad x_2=x_2(q_1,q_2,q_3),\quad x_3=x_3(q_1,q_2,q_3).
\end{equation}

We carry out the change of variables by multiplying the unites
\begin{equation}
\label{Di3d}
\mathbbm{1}_i=
\int\,dq_i\,\delta[q_i-q_i(x_1,x_2,x_3)]=1
\end{equation}
for $i=1$ through $3$ to $I_3$, sequentially.
First, after multiplying $\mathbbm{1}_1$ in Eq.~\eqref{Di3d} to
$I_3$, we find that $I_3$ can be expressed as
\begin{equation}
\label{I3d2}
I_3 = \int \,d q_1 
\int_{-\infty}^\infty\,dx_1\,dx_2\,dx_3\,
\delta[q_1-q_1(x_1,x_2,x_3)]
f(x_1,x_2,x_3).
\end{equation}

Similarly to the $2$-dimensional case, we perform
the integration over $x_1$ by making use of the Dirac delta function
for $q_1$ as
\begin{eqnarray}
\label{I3x1}
\int_{-\infty}^\infty\,dx_1\,\delta[q_1-q_1(x_1,x_2,x_3)]
&=&
\frac{1}{\left|\dfrac{\partial q_1}{\partial x_1}\right|_{x_1=x_1(q_1,x_2,x_3)}}
\nonumber\\
&=&
\frac{1}{\mathscr{G}_1(q_1, x_2, x_3)},
\end{eqnarray}
which leads to
\begin{equation}
\label{I3afterx1}
I_3 = \int \,d q_1 
\int_{-\infty}^\infty\,dx_2\,dx_3\,
\frac{1}{\mathscr{G}_1(q_1, x_2, x_3)}
f(q_1,x_2,x_3).
\end{equation}
After the integration over $x_1$,
every $x_1$ in the integrand and remaining delta functions is replaced with $x_1(q_1,x_2,x_3)$ satisfying the condition that the argument of the Dirac delta function vanishes,
$q_1-q_1(x_1,x_2,x_3)=0$.
Then the delta function for $q_2$ in Eq.~\eqref{Di3d} can be expressed as
$\delta[q_2 - q_2(q_1,x_2,x_3)]$.

After multiplying $\mathbbm{1}_2$ in Eq.~\eqref{Di3d} to $I_3$ in Eq.~\eqref{I3afterx1}, the integration over $x_2$ can be carried out by making use
of the Dirac delta function for $q_2$ as
\begin{eqnarray}
\label{I3x2}
\int_{-\infty}^\infty\,dx_2\,\delta[q_2-q_2(q_1,x_2,x_3)]
&=&
\frac{1}{\left|\left(\dfrac{\partial q_2}{\partial x_2}\right)_{(1)}\right|_{x_2=x_2(q_1,q_2,x_3)}}
\nonumber\\
&=&
\frac{\mathscr{G}_1(q_1, q_2, x_3)}{\mathscr{G}_2(q_1, q_2, x_3)},
\end{eqnarray}
where the last equality comes from Eq.~\eqref{I3j2j1}.
Then, Eq.~\eqref{I3x2} is expressed as
\begin{equation}
\label{I3afterx2}
I_3 = \int \,d q_1\,d q_2 
\int_{-\infty}^\infty\,dx_3\,
\frac{1}{\mathscr{G}_1(q_1, q_2, x_3)}
\times
\frac{\mathscr{G}_1(q_1, q_2, x_3)}{\mathscr{G}_2(q_1, q_2, x_3)}
f(q_1,q_2,x_3).
\end{equation}
$x_2(q_1,q_2,x_3)$ is determined from the condition that
the argument of the Dirac delta function vanishes,
$q_2-q_2(q_1,x_2,x_3)=0$.
Substituting $x_2(q_1,q_2,x_3)$ into $x_1(q_1,x_2,x_3)$, we obtain
$x_1=x_1(q_1,q_2,x_3)$ and the argument of the Dirac delta function
for $q_3$ in Eq.\eqref{Di3d} is expressed as $\delta[q_3-q_3(q_1,q_2,x_3)]$.

Finally, after multiplying $\mathbbm{1}_3$ in Eq.~\eqref{Di3d} to $I_3$ in Eq.~\eqref{I3afterx2}, we integrate over $x_3$ by taking into account
the delta function for $q_3$ as
\begin{eqnarray}
\label{I3x3}
\int_{-\infty}^\infty\,dx_3\,\delta[q_3-q_3(q_1,q_2,x_3)]
&=&
\frac{1}{\left|\left(\dfrac{\partial q_3}{\partial x_3}\right)_{(2)}\right|_{x_3=x_3(q_1,q_2,q_3)}}
\nonumber\\
&=&
\frac{\mathscr{G}_2(q_1, q_2, q_3)}{\mathscr{G}_3(q_1, q_2, q_3)},
\end{eqnarray}
where the last equality comes from Eq.~\eqref{I3j3j2}.
After the integration over $x_3$, every $x_3$ in the integrand and
the right-hand sides of
Eqs.~\eqref{I3x1}, \eqref{I3x2} and \eqref{I3x3} is replaced with
$x_3(q_1,q_2,q_3)$ which is determined from the condition that
the argument of the delta function vanishes,
$q_3-q_3(q_1,q_2,x_3)=0$.
Substituting $x_3(q_1,q_2,q_3)$ into
$x_1(q_1,q_2,x_3)$ and $x_2(q_1,q_2,x_3)$, we can obtain
the expressions for 
$x_1=x_1(q_1,q_2,q_3)$ and $x_2=x_2(q_1,q_2,q_3)$, 
respectively.
Then, we can express the integrand $f(x_1,x_2,x_3)$
in terms of $q_1$, $q_2$ and $q_3$ and
the integral $I_3$ is expressed as
\begin{eqnarray}
I_3&=&
\int \,d q_1 \, d q_2\, d q_3\,
\frac{1}{\mathscr{G}_1}
\times
\frac{\mathscr{G}_1}{\mathscr{G}_2}
\times
\frac{\mathscr{G}_2}{\mathscr{G}_3}
f[x_1(q_1,q_2,q_3),x_2(q_1,q_2,q_3),x_3(q_1,q_2,q_3)]
\nonumber\\
&=&
\int \,d q_1 \, d q_2\, d q_3\,
\mathscr{J}_3\,
f[x_1(q_1,q_2,q_3),x_2(q_1,q_2,q_3),x_3(q_1,q_2,q_3)],
\label{j3final}
\end{eqnarray}
where the Jacobian for the change of variables
is identified as $\mathscr{J}=\mathscr{J}_3=1/\mathscr{G}_3$.
This completes the proof of the Jacobian for a $3$-dimensional coordinate transformation or
change of variables.

%============================================================
\subsection{$\textbf{\textit{n}}$-dimensional case}
%============================================================

In this subsection, we consider the $n$-dimensional integral $I_n$ for a function $f(x_1,\cdots,x_n)$ 
defined in Eq.~\eqref{In} by multiplying the unities $\mathbbm{1}_i$ in Eq.~\eqref{trivial} 
to $I_n$ in Eq.~\eqref{In} sequentially.
Then, we integrate $I_n$, which is multiplied by the unity, 
over $x_i$ for $i=1$ through $n$ successively by making use of the Dirac delta function
\begin{equation}
\int_{-\infty}^\infty\,dx_i\,\delta[q_i-q_i(x_1,\cdots,x_n)].
\end{equation}
After the integration of all $x_i$ variables, the corresponding Jacobian formula is obtained by employing mathematical induction.

First, the integration over $x_1$ can be carried out from Eq.~\eqref{Ind} and
the result  for the integration over $x_1$ can easily be generalized from
the 2-dimensional version in
Eq.~\eqref{x1d2} as
\begin{eqnarray}
\label{x1dn}
\int_{-\infty}^\infty\,dx_1\,\delta[q_1-q_1(x_1,\cdots,x_n)]
&=&
\frac{1}{\left|
\dfrac{\partial q_1}{\partial x_1}
\right|_{x_1=x_1(q_1,x_2,\cdots,x_n)}}
\nonumber\\
&=&\frac{1}{\mathscr{G}_1(q_1, x_2,\cdots,x_n)},
\end{eqnarray}
which leads to 
\begin{eqnarray}
\label{Inafterx1}
I_n=
\int\,dq_1\,\int_{-\infty}^\infty
\,dx_2\cdots dx_n
\frac{1}{\mathscr{G}_1(q_1, x_2,\cdots,x_n)}
f(q_1,x_2,\cdots,x_n).
\end{eqnarray}
After the $x_1$ integration, every
$x_1$  in the integrand of Eq.~\eqref{Ind} and  $\mathscr{G}_1$  in Eq.~\eqref{x1dn} is replaced with
\begin{equation}
\label{xjafteri}
x_1=x_1(q_1,x_2,\cdots,x_n).
\end{equation}
The constraint equation coming from the convolution with the Dirac delta function
in Eq.~\eqref{x1dn} is
\begin{equation}
\label{q1xi}
q_1-q_1[x_1(q_1,x_2,\cdots,x_n),x_2,\cdots,x_n]=0.
\end{equation}

Then, we carry out the integration over $x_i$ for $i=1$ through $n-1$ by multiplying
$\mathbbm{1}_i$ to $I_n$ in Eq.~\eqref{Inafterx1} sequentially.
We assume that,
after the integration over  $x_i$ for $i=1$ through $n-1$, 
the result of the integration of Dirac delta functions is 
\begin{eqnarray}
\label{delta-integrated-j}
 \int_{-\infty}^{\infty}\, d x_1\cdots dx_i
\,
\prod_{j=1}^i\delta[q_j-q_j(x_1,\cdots,x_n)] 
&=&
\frac{1}{\mathscr{G}_1}
\times
\frac{\mathscr{G}_1}{\mathscr{G}_2}
\times \cdots \times
\frac{\mathscr{G}_{i-1}}{\mathscr{G}_i}
\nonumber\\
&=&
\frac{1}{\mathscr{G}_i(q_1,\cdots,q_i,x_{i+1},\cdots,x_n)}, 
\end{eqnarray}
which leads to
\begin{equation}
\label{Inafterxi}
I_n = \int \, d q_1\cdots dq_i
\int_{-\infty}^{\infty}\, d x_{i+1}\cdots dx_n
\frac{1}{\mathscr{G}_i(q_1,\cdots,q_i,x_{i+1},\cdots,x_n)}
f(q_1,\cdots,q_i,x_{i+1},\cdots,x_n).
\end{equation}
For any $j\le i$ every
$x_j$  in the integrand of Eq.~\eqref{In} and  $\mathscr{G}_j$'s in Eq.~\eqref{delta-integrated-j}  
is replaced with
\begin{equation}
\label{xjafteri2}
x_j=x_j(q_1,\cdots,q_i,x_{i+1},\cdots,x_n)
\end{equation}
after the integrations over $x_1$ through $x_i$. 
There are $i$ constraint equations coming from the convolution with Dirac delta functions 
in Eq.~\eqref{delta-integrated-j}:
\begin{equation}
\label{qjxi}
q_j-q_j[x_1(q_1,\cdots,q_i,x_{i+1},\cdots,x_n),\cdots,x_i(q_1,\cdots,q_i,x_{i+1},\cdots,x_n),x_{i+1},\cdots,x_n]=0,
\end{equation}
where $j$ runs from 1 through $i$.

After multiplying $\mathbbm{1}_{i+1}$ to $I_n$ in Eq.~\eqref{Inafterxi}, we integrate out one more Cartesian coordinate $x_{i+1}$ to find that
\begin{eqnarray}
\label{dxi1-int}
&&
\int_{-\infty}^\infty\,dx_{i+1}
\delta[q_{i+1}-q_{i+1}(q_1,\cdots,q_i,x_{i+1},\cdots,x_n)]_{x_j=x_j(q_1,\cdots,q_i,x_{i+1},\cdots,x_n)}
\nonumber\\
&=&\frac{1}{\left|\left(\dfrac{\partial q_{i+1}}{\partial x_{i+1}}\right)_{(i)}\right|_{x_k=x_k(q_1,\cdots,q_{i+1},x_{i+2},\cdots,x_n)}}
\nonumber \\
&=&
\frac{\mathscr{G}_{i}(q_1,\cdots,q_{i+1},x_{i+2},\cdots,x_n)}{\mathscr{G}_{i+1}(q_1,\cdots,q_{i+1},x_{i+2},\cdots,x_n)},
\end{eqnarray}
where $1\leq j\leq i$ and $1\leq k\leq i+1$. 
The proof of  the last equality can be found in
Eq.~\eqref{NDJacobianResult2} in Appendix \ref{APC}. 
In combination with Eqs.~\eqref{delta-integrated-j} and \eqref{dxi1-int}, we find that
\begin{eqnarray}
\int_{-\infty}^{\infty} d x_1\cdots dx_i\, dx_{i+1}\,
\,
\prod_{j=1}^{i+1}\delta[q_j-q_j(x_1,\cdots,x_n)] 
&=&
\frac{1}{\mathscr{G}_1}
\times
\frac{\mathscr{G}_1}{\mathscr{G}_2}
\times \cdots \times
\frac{\mathscr{G}_{i-1}}{\mathscr{G}_i}
\times
\frac{\mathscr{G}_{i}}{\mathscr{G}_{i+1}}
\nonumber \\
&=&
\frac{1}{\mathscr{G}_{i+1}(q_1,\cdots,q_{i+1},x_{i+2},\cdots,x_n)}.
\label{delta-integrated-jp1}%
\end{eqnarray}
For any $j\le i+1$ every
$x_j$  in the integrand of Eq.~\eqref{Ind} and  $\mathscr{G}_j$'s in Eq.~\eqref{delta-integrated-jp1} is
 replaced with
\begin{equation}
x_j=x_j(q_1,\cdots,q_{i+1},x_{i+2},\cdots,x_n)
\end{equation}
after the integrations over $x_1$ through $x_{i+1}$. 
There are $i+1$ constraint equations coming from the convolution with Dirac delta functions 
in Eq.~\eqref{delta-integrated-jp1}:
\begin{equation}
q_j-q_j[x_1(q_1,\cdots,q_{i+1},x_{i+2},\cdots,x_n),\cdots,x_i(q_1,\cdots,q_{i+1},x_{i+2},\cdots,x_n),x_{i+1},\cdots,x_n]=0,
\end{equation}
where $j$ runs from 1 through $i+1$.
According to mathematical induction,
this proves that the assumption in Eq.~\eqref{delta-integrated-j}
with the constraints \eqref{xjafteri2} and \eqref{qjxi}
is true for all $i=1$ through $n$.

Finally, the integral $I_n$ can be expressed in terms of $q_1$, $\cdots$, $q_n$ as
\begin{eqnarray}
I_n
&=&\int d^n{q}\,
\frac{1}{\mathscr{G}_n}
f(x_1,\cdots,x_n)|_{x_{k}=x_{k}(q_1,\cdots,q_n)}
\nonumber\\
&=&\int d^n{q}\,\mathscr{J}_n \, f(x_1,\cdots,x_n)|_{x_{k}=x_{k}(q_1,\cdots,q_n)},
\label{jnfinal}
\end{eqnarray}
where $1\leq k\leq n $. This completes the proof of the Jacobian $\mathscr{J}=\mathscr{J}_n=1/\mathscr{G}_n$ for an $n$-dimensional coordinate transformation or change of variables.

%%%%%%%%%%%%%%%%%%%%%%%%%%%%%%%%%%%%%%%%%%%%%%%%%%%%%%%%%%%%%
\section{Application \label{sec;app}}
%%%%%%%%%%%%%%%%%%%%%%%%%%%%%%%%%%%%%%%%%%%%%%%%%%%%%%%%%%%%%

Since the proof of the Jacobian formula in the previous section is rather abstract,
readers who are not familiar with the notation might be confused.
In this section, we present a few explicit examples of deriving the Jacobian
without resorting to the general formula derived in the previous section.
We expect that the explicit examples will help
readers to understand the method presented
in the previous section more intuitively
and to apply it to a specific change of variables.

%============================================================
\subsection{Spherical coordinates}
%============================================================

In this subsection, we consider the change of variables
from the $3$-dimensional Cartesian coordinates $(x,y,z)$ to the spherical coordinates $(r,\theta,\phi)$.
The Cartesian coordinates can be expressed in terms of
the radius $r$,
the polar angle $\theta$ and the azimuthal angle $\phi$
as
\begin{equation}
\label{xyz}
x=r\sin\theta\cos\phi,\quad y=r\sin\theta\sin\phi,\quad z=r\cos\theta.
\end{equation}
We use $\cos\theta$ instead of $\theta$ 
as the integration variable and
reorganize the order of multiple integrations in order to simplify the computation.
That is, $(x_1,x_2,x_3)$ in section \ref{3dcase} corresponds to $(z,y,x)$ while
$(q_1,q_2,q_3)$ corresponds to $(\cos\theta, \phi, r)$, respectively.
However, it turns out that the integral is invariant under this reordering.
We also note that one can use, for instance, $\sin\theta$ instead of
$\cos\theta$ as an integration variable and it will not
alter the result of the integration.
Then the inverse transformation of Eq.~\eqref{xyz} is expressed as
\begin{subequations}
\begin{eqnarray}
\cos \theta &=&
\frac{z}{\sqrt{x^2+y^2+z^2}}, 
\\
\phi &=&
\delta \left(
\phi - \arctan\frac{y}{x}
\right)
\Theta(x)
+
\delta \left(
\phi - \arctan\frac{y}{x}-\pi
\right)
\Theta(-x),
\label{thetaphi}
\\
r&=&\sqrt{x^2+y^2+z^2},
\end{eqnarray}
\end{subequations}
where the Heaviside step function $\Theta(x)$ is
defined by
\begin{equation}
\Theta(x)=\begin{cases}
1,&\textrm{for}\,\,x>0,
\\
\frac{1}{2}, &\textrm{for}\,\, x=0,
\\
0,&\textrm{for}\,\, x<0.
\end{cases}
\end{equation}
Conventionally, the arctangent function is defined in the region $[-\frac{\pi}{2},\frac{\pi}{2}]$.
The period of the tangent function is $\pi$, while the azimuthal angle ranges
from 0 to $2\pi$. Thus the angle $\phi$ for $x>0$ is set to be $\arctan\frac{y}{x}\in[-\frac{\pi}{2},\frac{\pi}{2}]$
while that for $x<0$ is set to be $\pi+\arctan\frac{y}{x}\in[\frac{\pi}{2},\frac{3\pi}{2}]$
in order to make the transformation function invertible
in the entire range:
\begin{equation}
\phi=
\begin{cases}
\arctan\frac{y}{x}\in(-\frac{\pi}{2},\frac{\pi}{2}),& \textrm{for}\,\, x>0,\\
\pi+\arctan\frac{y}{x}\in( \frac{\pi}{2},\frac{3\pi}{2}),&\textrm{for}\,\,  x<0,\\
\frac{\pi}{2},& \textrm{for}\,\, x=0\,\,\textrm{and}\,\, y>0,\\
-\frac{\pi}{2},& \textrm{for}\,\, x=0\,\,\textrm{and}\,\, y<0,\\
0,&\textrm{for}\,\,x=y=0.
\end{cases}
\end{equation}

We consider a $3$-dimensional integral $J_3$ with an arbitrary integrand
 $f(x,y,z)$
\begin{equation}
\label{J3}
J_3=\int_{-\infty}^\infty dx \, dy \, dz \,f(x,y,z).
\end{equation}
We multiply the unities
\begin{subequations}
\begin{eqnarray}
\label{th3}
\mathbbm{1}_1&=&
\int_{-1}^1d\cos\theta\,
\delta\left(\cos\theta-\frac{z}{\sqrt{x^2+y^2+z^2}}\right)=1,
\\
\label{phi3}
\mathbbm{1}_2&=&
\int_{-\pi/2}^{3\pi/2}d\phi\,
\left[
\delta
\left(
\phi-\arctan\frac{y}{x}
\right)\Theta(x)
+\delta
\left(\phi-\arctan\frac{y}{x}-\pi\right)\Theta(-x)
\right]=1,
\\
\label{r3}
\mathbbm{1}_3&=&
\int_0^\infty  dr \,
\delta\left(r-\sqrt{x^2+y^2+z^2}\right)=1
\end{eqnarray}
\end{subequations}
to $J_3$ without changing the value of the integral sequentially.

First, we integrate out the $x_1=z$ coordinate by multiplying $\mathbbm{1}_1$ in Eq. \eqref{th3} to $J_3$ in Eq. \eqref{J3}.
The integration over $z$ can be performed as
\begin{equation}
\int_{-\infty}^{\infty} dz\,\delta\left(\cos\theta-\tfrac{\displaystyle z}{\sqrt{\displaystyle x^2+y^2+z^2}}\right)=\frac{\sqrt{x^2+y^2}}{\sin^3\theta}.
\label{J3th}
\end{equation}
After the integration over $z$, every $z$ in the integrand of $J_3$ is replaced 
with the expression in terms of $\cos\theta$, $y$ and $x$:
 $z=\sqrt{x^2+y^2}/\tan\theta$, where
 we omit the simple conversion between trigonometric functions here and after.
The integrand of the remaining double integral over $x$ and $y$ is a function of 
$\theta$, $x$ and $y$:
\begin{equation}
\label{J3Az}
J_3=\int_{-1}^1d\cos\theta\,
\int_{-\infty}^\infty dx \, dy \,f(x,y,\sqrt{x^2+y^2}/\tan\theta)\frac{\sqrt{x^2+y^2}}{\sin^3\theta}.
\end{equation}
This can also be obtained by substituting $\cos\theta$, $y$ and $x$ into Eq.~\eqref{I3x1}.

Next, we integrate out the $x_2=y$ coordinate by multiplying $\mathbbm{1}_2$ in Eq. \eqref{phi3} into $J_3$ in Eq. \eqref{J3Az}.
The integration over $y$ can be performed as
\begin{eqnarray}
\int_{-\infty}^\infty \,dy \left[
\delta\left(\phi-\arctan\frac{y}{x}\right)\Theta(x)
+
\delta\left(\phi-\arctan\frac{y}{x}-\pi\right)\Theta(-x)
\right]
&=&\frac{1}{|\frac{d}{dy}\arctan\frac{y}{x}|}[\Theta(x)+\Theta(-x)]\Big|_{y=x\tan\phi}
\nonumber\\
%&=&\frac{x^2(1+\tan^2\phi)}{|x|}
%\nonumber\\
\label{J3phi}
&=&\frac{|x|}{\cos^2\phi}
,
\end{eqnarray}
where 
we have used the identity
\begin{eqnarray}
\Theta(x)+\Theta(-x)&=&1.
\end{eqnarray}
Equation~\eqref{J3phi} can also be obtained by substituting $\cos\theta$, $\phi$ and $x$ into Eq.~\eqref{I3x2}. After the integration over $y$, every $y$ in the integrand of $J_3$ is replaced with
the expression in terms of $\cos\theta$, $\phi$ and $x$. 
The integrand of the remaining integral over $x$ is a function of $\cos\theta$, $\phi$ and $x$:
\begin{eqnarray}
J_3
&=&\int_{-1}^1d\cos\theta\,\int_{0}^{2\pi}d\phi\,
\int_{-\infty}^\infty dx  \,f(x,
y
,\sqrt{x^2+y^2}/\tan\theta)\frac{\sqrt{x^2+y^2}}{\sin^3\theta}\frac{|x|}{\cos^2\phi}\Bigg|_{y=x\tan\phi}
\nonumber\\
&=&\int_{-1}^1d\cos\theta\,\int_{0}^{2\pi}d\phi\,
\int_{-\infty}^\infty dx  \,f[x,x\tan\phi,|x|/(|\cos\phi|\tan\theta)]
\frac{x^2}{\sin^3\theta|\cos\phi|^3},
\label{J3Aphi}%
\end{eqnarray}
where $y=x\tan\phi$.
After the integrations over both $z$ and $y$,
$y=x\tan\phi$ and $z=x/(\cos\phi\tan\theta)$.

Finally, after multiplying $\mathbbm{1}_3$ in Eq. \eqref{r3} into $J_3$ in Eq. \eqref{J3Aphi}, the integration over $x_3=x$ can be performed as
\begin{eqnarray}
\int_{-\infty}^\infty dx\,\delta(r-\sqrt{x^2+y^2+z^2})\Big|_{y=x\tan\phi,z=\frac{x}{\cos\phi\tan\theta}} 
&=&\int_{-\infty}^\infty dx\,\delta\left[r-\left|\frac{x}{\cos\phi\sin\theta}\right|\right]  
\nonumber\\
\label{J3r}
&=&|\cos\phi|\sin\theta.
\end{eqnarray}
This can also be obtained by substituting $\cos\theta$, $\phi$ and $r$ into Eq.~\eqref{I3x3}. Here, the radius $r$ is non-negative because
the Dirac delta function requires that 
$r=\sqrt{x^2+y^2+z^2}$ and $x^2+y^2+z^2$ is non-negative.
After the integration over all of the Cartesian coordinates, $x$, $y$ and $z$ are expressed
as in Eq.~\eqref{xyz}.

Substituting Eq.~\eqref{xyz} into 
\eqref{J3th}, \eqref{J3phi}, \eqref{J3r},
and $f(x,y,z)$,
we find that $J_3$ can be expressed in terms of the spherical coordinates as
\begin{eqnarray}
J_3
&=&\int_0^\infty  dr   
 \int_{-1}^{1}d\cos\theta
 \int_0^{2\pi}d\phi\,\frac{r\sin\theta}{\sin^3\theta}\frac{|r\sin\theta\cos\phi|}{\cos^2\phi}|\cos\phi|\sin\theta
\,\,f(r\sin\theta\cos\phi,r\sin\theta\sin\phi,r\cos\theta)
\nonumber\\
&=&\int_0^\infty  dr   
 \int_{-1}^{1}d\cos\theta
 \int_0^{2\pi}d\phi\,\,r^2f(r\sin\theta\cos\phi,r\sin\theta\sin\phi,r\cos\theta)
 \nonumber\\
&=&\int_0^\infty  dr   
 \int_{0}^{\pi}d\theta
 \int_0^{2\pi}d\phi\,\,r^2\sin\theta\,\,f(r\sin\theta\cos\phi,r\sin\theta\sin\phi,r\cos\theta)
 .
\end{eqnarray}
The overall factor $r^2\sin\theta$ is identified to be the Jacobian
\begin{equation}
\mathscr{J}=r^2\sin\theta.
\end{equation}
This exactly reproduces the result which can be obtained 
by applying the formula \eqref{j3final}~\cite{arfken}.

%============================================================
\subsection{$\textbf{\textit{n}}$-dimensional polar coordinates}
%============================================================

In this subsection, we extend the $3$-dimensional case
to the $n$-dimensional coordinate transformation from
the $n$-dimensional Cartesian coordinates
$(x_1,\cdots,x_n)$ to the $n$-dimensional polar coordinates
$(r,\theta_1,\theta_2,\cdots,\theta_{n-2},\phi)$.
Here,  $r$ is the radius and
there are $n-2$ polar angles $\theta_i$'s and a single azimuthal angle $\phi$.
The corresponding transformation functions are expressed as
\begin{eqnarray}
x_1&=&r\sin\theta_1\sin\theta_2\cdots\sin\theta_{n-3}\sin\theta_{n-2}\cos\phi,
\nonumber\\
x_2&=&r\sin\theta_1\sin\theta_2\cdots\sin\theta_{n-3}\sin\theta_{n-2}\sin\phi,
\nonumber\\
x_3&=&r\sin\theta_1\sin\theta_2\cdots\sin\theta_{n-3}\cos\theta_{n-2},
\nonumber\\
&\vdots&
\nonumber\\
x_{n-1}&=&r\sin\theta_1\cos\theta_2,
\nonumber\\
x_n&=&r\cos\theta_1.
\label{x1xn}
\end{eqnarray}
We reorganize the order of integrations of the Cartesian coordinates as $(x_n,x_{n-1},\cdots,x_1)$
for simplicity and the corresponding curvilinear coordinates are reorganized as $(\cos\theta_1,\cos\theta_2,\cdots,\cos\theta_{n-2},\phi,r)$.
We consider an $n$-dimensional integral $J_n$
with an arbitrary integrand $f(x_1,x_2,\cdots,x_n)$:
\begin{equation}
J_n=\int_{-\infty}^\infty d^n x \,f(x_1,x_2,\cdots,x_n),
\end{equation}
where $d^n x=dx_1 dx_2 \cdots dx_n$.
We multiply the unities
\begin{subequations}
\begin{eqnarray}
\mathbbm{1}_i&=&\int_{-1}^1
d\cos\theta_i
\,\delta\left(
\cos\theta_i-\frac{x_{n-i+1}}{\sqrt{x_1^2+\cdots+x_{n-i+1}^2}}\right)=1,\quad i=1,\,\cdots,n-2,
\label{costi}
\\
\mathbbm{1}_{n-1}&=&\int_{-\frac{\pi}{2}}^{\frac{3}{2}\pi}d\phi
\left[
\delta
\left(\phi-\arctan\frac{x_2}{x_1}\right)\Theta(x_1)
+\delta
\left(\phi-\arctan\frac{x_2}{x_1}-\pi\right)\Theta(-x_1)
\right]=1,
\label{thetaphin}
\\
\mathbbm{1}_n&=&
\int_0^\infty  dr\,
\delta\left(r-\sqrt{x_1^2+\cdots+x_n^2}\right)=1
\label{r-int}
\end{eqnarray}
\end{subequations}
for $i=1$ through $n$
to $J_n$, sequentially, keeping the integral invariant.
$\Theta(x)$ in Eq.~\eqref{thetaphin} is the Heaviside step function defined in Eq.~\eqref{thetaphi}.
There are numerous ways to perform the multiple integrations over the $n$ Cartesian coordinates. 
Our strategy to integrate out $x_i$'s is as follows:
According to the integrand of the right-hand side of Eq.~\eqref{costi}, the integration over
$x_{n-i+1}$ in the integral $J_n$ provides the constraint to the polar angle $\theta_i$.
Thus we choose to integrate over from $x_n$ to $x_3$ to express them in terms of
the polar angles from $\theta_1$ through $\theta_{n-2}$. 
Then we integrate out $x_2$ to express $x_n$ through $x_2$ in terms of
the $n-2$ polar angles and the azimuthal angle $\phi$ by making use
of Eq.~\eqref{thetaphin}. As the last step, we integrate out $x_1$ to determine
all of the Cartesian coordinates in terms of the spherical polar coordinates by
making use of \eqref{r-int}.

First, after multiplying $\mathbbm{1}_i$ in Eq.~\eqref{costi} into $J_n$, the integration over $x_{n-i+1}$ for $i=1$ through $n-2$  can be performed as
\begin{equation}
\label{thintegration}
\int_{-\infty}^\infty dx_{n-i+1}
\,\delta
\left(
\cos\theta_i-\frac{x_{n-i+1}}{\sqrt{x_1^2+\cdots+x_{n-i+1}^2}}
\right)
=\frac{\sqrt{x_1^2+\cdots+x_{n-i}^2}}{\sin^3\theta_i},
\end{equation}
where one can obtain the same results from Eqs.~\eqref{x1dn} and \eqref{dxi1-int} taking care of
the order of integration.
After the integration over $x_{n-i+1}$, we make the replacement
\begin{equation}
\label{xn-i+1}
x_{n-i+1}=\frac{\sqrt{x_1^2+\cdots+x_{n-i}^2}}{\tan\theta_i}.
\end{equation}
Then, $J_n$ is expressed as
\begin{equation}
\label{Jnafterxn-2}
J_n = 
\int_{-1}^1 \, d\cos\theta_1 \cdots \int_{-1}^1 \,d\cos\theta_{n-2}
\,
\int_{-\infty}^\infty dx_1 dx_2
f(x_1,x_2,\cos\theta_1,\cdots,\cos\theta_{n-2})
\,
\prod_{i=1}^{n-2}
\frac{\sqrt{x_1^2+\cdots+x_{n-i}^2}}{\sin^3\theta_i},
\end{equation}
where every $x_i$ in the last factor for $i=3$ through $n$ is replaced by that in Eq.~\eqref{xn-i+1}.

After multiplying $\mathbbm{1}_{n-1}$ in Eq.~\eqref{thetaphin} into $J_n$ in Eq.~\eqref{Jnafterxn-2}, the integration over $x_2$ can be carried out 
in a similar manner as is done in Eq.~\eqref{J3phi}. The result is
\begin{eqnarray}
\int_{-\infty}^\infty dx_2
\,\left[
\delta\left(\phi-\arctan\frac{x_2}{x_{1}}\right)\Theta(x_1)
+
\delta\left(\phi-\arctan\frac{x_2}{x_1}-\pi\right)\Theta(-x_1)
\right]
&=&\frac{|x_1|}{\cos^2\phi}
,
\label{x2integration}
\end{eqnarray}
where 
$x_2=x_1\tan\phi$ after the integration.
This can also be obtained from Eq.~\eqref{dxi1-int} while keeping the results in Eqs.~\eqref{thintegration} and \eqref{xn-i+1}.
Then, $x_{n-i+1}$ for $i=1$ through $n-1$ can be expressed as
\begin{equation}
\label{xni1}
x_{n-i+1}=\frac{|x_1|}{|\cos\phi|\sin\theta_{n-2}\cdots\sin\theta_{i+1}\tan\theta_i}.
\end{equation}
Then, we find that
\begin{equation}
\label{Jnafterxn-1}
J_n = 
\int_{-1}^1 \, d\cos\theta_1 \cdots \int_{-1}^1 \,d\cos\theta_{n-2}
\int_{0}^{2\pi} \,d\phi
\,
\int_{-\infty}^\infty dx_1 
f(x_1,\phi,\cos\theta_1,\cdots,\cos\theta_{n-2})
\,\frac{|x_1|}{\cos^2\phi}
\,
\prod_{i=1}^{n-2}
\frac{\sqrt{x_1^2+\cdots+x_{n-i}^2}}{\sin^3\theta_i},
\end{equation}
where every $x_i$ in the last factor for $i=2$ through $n$ is replaced by that in Eq.~\eqref{xni1}.

Finally, after multiplying $\mathbbm{1}_n$ in Eq. \eqref{r-int}
to $J_n$ in Eq.~\eqref{Jnafterxn-1}, the integration over $x_1$ can be performed like Eq.~\eqref{J3r} and we obtain
\begin{eqnarray}
\int_{-\infty}^\infty dx_1 \,\delta\left(r-\sqrt{x_1^2+\cdots+x_n^2}\right)
&=&\int_{-\infty}^\infty dx_1\,\delta\left[r- \left|\frac{x_1}{\cos\phi\sin\theta_{n-2}\cdots\sin\theta_1}\right|\right]  
\nonumber\\
&=&\left|\cos\phi\right|\sin\theta_{n-2}\cdots\sin\theta_1,
\end{eqnarray}
where we have omitted the replacements
of 
$x_2=x_1\tan\phi$ and $x_{3}$ through $x_n$
that can be obtained from Eq.~\eqref{xni1}
on the left-hand side.
This results can also be obtained from Eq.~\eqref{dxi1-int} while keeping the results in Eqs.~\eqref{thintegration}, \eqref{xn-i+1}, \eqref{x2integration} and \eqref{xni1}.
After integrating out all of the Cartesian coordinates, 
we reproduce the expression for every $x_i$ that is given in Eq.~\eqref{x1xn}.

Combining all of the results listed above, we find that 
the $n$-dimensional coordinate transformation or
change of variables is carried out as
\begin{eqnarray}
J_n&=&
\int_0^\infty  dr   
 \int_{-1}^{1}d\cos\theta_1
 \cdots
 \int_{-1}^{1}d\cos\theta_{n-2}
 \int_0^{2\pi}d\phi
 \nonumber\\
 &&\times
\frac{r\sin\theta_1}{\sin^3\theta_1}\cdots\frac{r\sin\theta_1\cdots\sin\theta_{n-2}}{\sin^3\theta_{n-2}}\frac{r\sin\theta_1\cdots\sin\theta_{n-2}|\cos\phi|}{\cos^2\phi}|\cos\phi|\sin\theta_1\cdots\sin\theta_{n-2}
\nonumber\\
&&\times f(r\sin\theta_1\cdots\sin\theta_{n-2}\cos\phi,r\sin\theta_1\cdots\sin\theta_{n-2}\sin\phi,r\sin\theta_1\cdots\sin\theta_{n-3}\cos\theta_{n-2},\cdots,r\sin\theta_1\cos\theta_2,r\cos\theta_1)
\nonumber\\
&=&\int_0^\infty  dr   
 \int_{0}^{\pi}d\theta_1
 \cdots
 \int_{0}^{\pi}d\theta_{n-2}
 \int_0^{2\pi}d\phi\,\,
r^{n-1}\sin^{n-2}\theta_1\cdots\sin^2\theta_{n-3}\sin\theta_{n-2}
\nonumber\\
&&\times f(r\sin\theta_1\cdots\sin\theta_{n-2}\cos\phi,r\sin\theta_1\cdots\sin\theta_{n-2}\sin\phi,r\sin\theta_1\cdots\sin\theta_{n-3}\cos\theta_{n-2},\cdots,r\sin\theta_1\cos\theta_2,r\cos\theta_1).
\nonumber
\\
\end{eqnarray}
The extra factor $r^{n-1}\sin^{n-2}\theta_1\cdots\sin^2\theta_{n-3}\sin\theta_{n-2}$ 
in front of the original integrand $f$
is identified with
 the Jacobian
\begin{eqnarray}
\mathscr{J}=r^{n-1}\sin^{n-2}\theta_1\cdots\sin^2\theta_{n-3}\sin\theta_{n-2}.
\end{eqnarray}
This exactly reproduces the result in Refs.~\cite{nsphere,combinatorics}, 
 which can be obtained 
by applying the general formula \eqref{jnfinal}.

%============================================================
\section{Conclusions\label{sec;con}}
%============================================================

We have derived the general formula for the Jacobian
of the transformation from the $n$-dimensional 
Cartesian coordinates to arbitrary curvilinear coordinates
by making use of Dirac delta functions, whose
arguments correspond to the transformation functions between the two coordinate systems.
The multiplication of the trivial identities \eqref{trivial}
to the original integral enables us to
integrate out the original integration variables
corresponding to the Cartesian coordinates systematically.
By making use of the chain rule for the partial derivatives,
we can carry out the integration over the Cartesian 
coordinates successively and end up with the integral
expressed in terms of the curvilinear coordinates.
Then, the Jacobian can be read off by comparing the integrands
of the resultant integral with the original one.
It turns out that the formula derived in this paper
exactly reproduces the Jacobian for the coordinate
transformation or change of variables.

We have presented a few examples, where we have integrated 
out the Cartesian coordinates by making use of Dirac delta functions explicitly without resorting to the general formula
for the Jacobian
derived in this paper.
We find that the formulas obtained in these explicit examples
are exactly the same as those in the general formula \eqref{jnfinal}.
Since the derivation of the Jacobian in the general case that makes use of
the chain rule of the partial derivatives
is rather abstract, we expect that these examples
will give insights on understanding 
the derivation concretely.

To our best knowledge, this derivation of the Jacobian
factor by making use of Dirac delta functions
for the coordinate transformation or change of variables
from the $n$-dimensional Cartesian coordinates
to the curvilinear coordinates is new.
Although there are several ways to derive the Jacobian available 
in textbooks~\cite{arfken,stewart,apostol}, our derivation could be pedagogically useful
in upper-level mathematics or physics courses
in practice using Dirac delta functions successively.
Compared with the methods popular in the textbook level,
our method is more intuitive because we have
employed only the explicit calculation of elementary single-dimensional integrals
without relying on abstract geometrical interpretations or
more abstract Green's theorem with which undergraduate 
physics-major students are not usually familiar.
Furthermore, a detailed derivation of the chain rule for the partial derivatives, 
which is employed to prove the Jacobian formula,  
should be a nice working example with which one can understand 
a rigorous usage of the partial derivatives with multi-dimensional variables
without ambiguity.

\acknowledgments
As members of the Korea Pragmatist Organization for Physics Education
(\textsl{KPOP}$\mathscr{E}$), 
the authors thank the remaining members of \textsl{KPOP}$\mathscr{E}$ 
for useful discussions.
This work is supported in part by the National Research Foundation of Korea (NRF) under the BK21 FOUR program at Korea University, Initiative for science frontiers on upcoming challenges,
and by grants funded by the Korea government (MSIT),
Grants No. NRF-2017R1E1A1A01074699 (J.L.)
and No. NRF-2020R1A2C3009918 (J.E. and D.K.).
The work of C.Y. is supported by Basic Science Research Program through the National Research Foundation (NRF) of Korea funded by the Ministry of Education (2020R1I1A1A01073770).

\appendix

\section{ $\textbf{2}$-dimensional case\label{APA}}

In this section, we consider a $2$-dimensional coordinate transformation from 
the Cartesian coordinates
$(x_1,x_2)$ to the curvilinear coordinates $(q_1,q_2)$.
The total differentials of $q_1$ and $x_1$ can be expressed as
\begin{subequations}
\begin{eqnarray}
dq_1&=&\frac{\partial q_1}{\partial x_1}dx_1+\frac{\partial q_1}{\partial x_2}dx_2,
\label{dq1}
\\
dx_1&=&\left(\frac{\partial x_1}{\partial q_1}\right)_{(1)}dq_1+\left(\frac{\partial x_1}{\partial x_2}\right)_{(1)}dx_2,
\label{dx1}%
\end{eqnarray}
\end{subequations}
where $q_1=q_1(x_1,x_2)$ in Eq.~\eqref{dq1}
and $x_1=x_1(q_1,x_2)$ in Eq.~\eqref{dx1}, respectively.
Note that
the definition of the partial derivative with a subscript 
are given in section \ref{subsec:SN}:
$\left({\partial x_1}/{\partial q_1}\right)_{(1)}$,
$\left({\partial x_1}/{\partial x_2}\right)_{(1)}$,
and $(\partial q_1 / \partial x_{1})_{(2)}$ are 
 defined in Eq.~\eqref{eq:pd-xq},
 \eqref{eq:pd-xx} and \eqref{eq:pd-qxx}, respectively.

Replacing $dx_1$ in Eq.~\eqref{dq1} with the right-hand side of Eq.~\eqref{dx1}, we obtain
\begin{eqnarray}
dq_1=\frac{\partial q_1}{\partial x_1}\left(\frac{\partial x_1}{\partial q_1}\right)_{(1)}dq_1+\left[\frac{\partial q_1}{\partial x_1}\left(\frac{\partial x_1}{\partial x_2}\right)_{(1)}+\frac{\partial q_1}{\partial x_2}\right]dx_2.
\end{eqnarray}
Comparing the coefficients of the differentials on both sides, we find that
\begin{subequations}
\begin{eqnarray}
\frac{\partial q_1}{\partial x_1}\left(\frac{\partial x_1}{\partial q_1}\right)_{(1)}&=&1,
\label{cr1}
\\
\frac{\partial q_1}{\partial x_1}\left(\frac{\partial x_1}{\partial x_2}\right)_{(1)}+\frac{\partial q_1}{\partial x_2}&=&0.
\label{cr2}%
\end{eqnarray}
\end{subequations}
We note that  Eq.~\eqref{cr1} is trivial. 
By making use of Eq.~\eqref{cr2}, we find that 
the factor in the denominator of Eq.~\eqref{x2d2} 
$\left( \partial q_2 / \partial x_2 \right)_{(1)}$ 
can be expressed as
\begin{equation}
\left(\dfrac{\partial q_2}{\partial x_2}\right)_{(1)}
=\frac{\partial q_2}{\partial x_1}\left(\frac{\partial x_1}{\partial x_2}\right)_{(1)}+\frac{\partial q_2}{\partial x_2}
%\frac{
%\dfrac{\partial q_1}{\partial x_1}
%\dfrac{\partial q_2}{\partial x_2}
%-\dfrac{\partial q_2}{\partial x_1}
%\dfrac{\partial q_1}{\partial x_2}
%}
%{\dfrac{\partial q_1}{\partial x_1}}
=
\frac{
\mathscr{Det}
\begin{pmatrix}
\dfrac{\partial q_1}{\partial x_1} &
\dfrac{\partial q_2}{\partial x_1}
\\
\dfrac{\partial q_1}{\partial x_2} &
\dfrac{\partial q_2}{\partial x_2}
\end{pmatrix}
}
{\mathscr{Det}\left(\dfrac{\partial q_1}{\partial x_1}\right)},
\end{equation}
%\cite{Blendell}
%\begin{eqnarray}
%\label{useofchainrule}
%\left(\frac{\partial x_1}{\partial x_2}\right)_{q_1}&=&-\frac{\left(\dfrac{\partial q_1}{\partial %x_2}\right)_{x_1}}{\left(\dfrac{\partial q_1}{\partial x_1}\right)_{x_2}},
%\nonumber\\
%\left(\frac{\partial q_2}{\partial x_2}\right)_{q_1}&=&
%\left(\frac{\partial q_2}{\partial x_1}\right)_{x_2}
%\left(\frac{\partial x_1}{\partial x_2}\right)_{q_1}
%+\left(\frac{\partial q_2}{\partial x_2}\right)_{x_1}
%=-\left(\frac{\partial q_2}{\partial x_1}\right)_{x_2}
%\frac{\left(\dfrac{\partial q_1}{\partial x_2}\right)_{x_1}}
%{\left(\dfrac{\partial q_1}{\partial x_1}\right)_{x_2}}
%+\left(\frac{\partial q_2}{\partial x_2}\right)_{x_1}.
%\end{eqnarray}
%Then, it is trivial to find the following relation
%\begin{equation}
%\left(\dfrac{\partial q_1}{\partial x_1}\right)_{x_2}
%\left(\dfrac{\partial q_2}{\partial x_2}\right)_{q_1}
%=
%\left(\dfrac{\partial q_1}{\partial x_1}\right)_{x_2}
%\left(\dfrac{\partial q_2}{\partial x_2}\right)_{x_1}
%-\left(\dfrac{\partial q_2}{\partial x_1}\right)_{x_2}
%\left(\dfrac{\partial q_1}{\partial x_2}\right)_{x_1}
%=\mathscr{Det}
%\begin{pmatrix}
%\dfrac{\partial q_1}{\partial x_1} &
%\dfrac{\partial q_2}{\partial x_1}
%\\
%\dfrac{\partial q_1}{\partial x_2} &
%\dfrac{\partial q_2}{\partial x_2}
%\end{pmatrix}
%\end{equation}
which leads to 
\begin{equation}
\label{j2j1}
\left|
\left(\dfrac{\partial q_2}{\partial x_2}\right)_{(1)}
\right|
=\frac{\mathscr{G}_2}{\mathscr{G}_1}.
\end{equation}

\section{$\textbf{3}$-dimensional case\label{APB}}

In this section, we consider a $3$-dimensional coordinate transformation from 
the Cartesian coordinates
$(x_1,x_2,x_3)$ to the curvilinear coordinates $(q_1,q_2,q_3)$.
First, we consider the total differentials of $q_1$ and $x_1$ which can be expressed as
\begin{subequations}
\begin{eqnarray}
\label{I3dq1}
dq_1&=&\frac{\partial q_1}{\partial x_1}dx_1+\frac{\partial q_1}{\partial x_2}dx_2+\frac{\partial q_1}{\partial x_3}dx_3,
\\
\label{I3dx1}
dx_1&=&\left(\frac{\partial x_1}{\partial q_1}\right)_{(1)}dq_1+\left(\frac{\partial x_1}{\partial x_2}\right)_{(1)}dx_2+\left(\frac{\partial x_1}{\partial x_3}\right)_{(1)}dx_3,
\end{eqnarray}
\end{subequations}
where $q_1=q_1(x_1,x_2,x_3)$ in Eq.~\eqref{I3dq1}
and $x_1=x_1(q_1,x_2,x_3)$ in Eq.~\eqref{I3dx1}, respectively.
Replacing $d x_1$ in Eq.~\eqref{I3dq1} with the right-hand side
of Eq.~\eqref{I3dx1}, we obtain
\begin{eqnarray}
dq_1
%&=&\frac{\partial q_1}{\partial x_1}\left[\left(\frac{\partial x_1}{\partial q_1}\right)_{(1)}dq_1+\left(\frac{\partial x_1}{\partial x_2}\right)_{(1)}dx_2+\left(\frac{\partial x_1}{\partial x_3}\right)_{(1)}dx_3\right]+\frac{\partial q_1}{\partial x_2}dx_2+\frac{\partial q_1}{\partial x_3}dx_3
%\nonumber\\
&=&\frac{\partial q_1}{\partial x_1}\left(\frac{\partial x_1}{\partial q_1}\right)_{(1)}dq_1+\left[\frac{\partial q_1}{\partial x_1}\left(\frac{\partial x_1}{\partial x_2}\right)_{(1)}+\frac{\partial q_1}{\partial x_2}\right]dx_2+\left[\frac{\partial q_1}{\partial x_1}\left(\frac{\partial x_1}{\partial x_3}\right)_{(1)}+\frac{\partial q_1}{\partial x_3}\right]dx_3.
\nonumber\\
\end{eqnarray}
Comparing the coefficients of the differentials on both sides, we find that
\begin{subequations}
\begin{eqnarray}
\label{c3r1}
\frac{\partial q_1}{\partial x_1}\left(\frac{\partial x_1}{\partial q_1}\right)_{(1)}&=&1,
\\
\label{c3r2}
\frac{\partial q_1}{\partial x_1}\left(\frac{\partial x_1}{\partial x_2}\right)_{(1)}+\frac{\partial q_1}{\partial x_2}&=&0,
\\
\label{c3r3}
\frac{\partial q_1}{\partial x_1}\left(\frac{\partial x_1}{\partial x_3}\right)_{(1)}+\frac{\partial q_1}{\partial x_3}&=&0,
\end{eqnarray}
\end{subequations}
where Eq.~\eqref{c3r1} is trivial.

By making use of Eq.~\eqref{c3r2}, we find that 
the factor in the denominator of Eq.~\eqref{I3x2} 
$\left( \partial q_2 / \partial x_2 \right)_{(1)}$ 
can be expressed as
\begin{equation}
\left(\frac{\partial q_2}{\partial x_2}\right)_{(1)}=\frac{\partial q_2}{\partial x_1}\left(\frac{\partial x_1}{\partial x_2}\right)_{(1)}+\frac{\partial q_2}{\partial x_2}
%=-\left(\frac{\partial q_2}{\partial %x_1}\right)_{x_2,x_3}\frac{\left(\dfrac{\partial %q_1}{\partial x_2}\right)_{x_1,x_3}}{\left(\dfrac{\partial %q_1}{\partial x_1}\right)_{x_2,x_3}}+\left(\frac{\partial %q_2}{\partial x_2}\right)_{x_1,x_3}
%=\frac{\dfrac{\partial q_1}{\partial x_1}\dfrac{\partial q_2}{\partial x_2}-\dfrac{\partial q_2}{\partial x_1}\dfrac{\partial q_1}{\partial x_2}}{\dfrac{\partial q_1}{\partial x_1}}
=
\frac{\mathscr{Det}
	\begin{pmatrix}
	\dfrac{\partial q_1}{\partial x_1} & \dfrac{\partial q_2}{\partial x_1}
	\\
	\dfrac{\partial q_1}{\partial x_2} & \dfrac{\partial q_2}{\partial x_2}
	\end{pmatrix}
}{\mathscr{Det}
\left(\dfrac{\partial q_1}{\partial x_1}\right)},
\end{equation}
which leads to
\begin{equation}
\label{I3j2j1}
\left|
\left(\dfrac{\partial q_2}{\partial x_2}\right)_{(1)}
\right|
=\frac{\mathscr{G}_2}{\mathscr{G}_1}.
\end{equation}

In order to prove Eq.~\eqref{I3j3j2}, we take into account
the total differentials of $q_1$, $x_1$, $q_2$ and $x_2$.
The total derivatives can be expressed as
\begin{subequations}
\begin{eqnarray}
\label{I3dq12}
dq_1&=&\dfrac{\partial q_1}{\partial x_1}dx_1+\dfrac{\partial q_1}{\partial x_2}dx_2+\dfrac{\partial q_1}{\partial x_3}dx_3,
\\
\label{I3dq2}
dq_2&=&\dfrac{\partial q_2}{\partial x_1}dx_1+\dfrac{\partial q_2}{\partial x_2}dx_2+\dfrac{\partial q_2}{\partial x_3}dx_3,
\\
\label{I3dx12}
dx_1&=&\left(\dfrac{\partial x_1}{\partial q_1}\right)_{(2)}dq_1+\left(\dfrac{\partial x_1}{\partial q_2}\right)_{(2)}dq_2+\left(\dfrac{\partial x_1}{\partial x_3}\right)_{(2)}dx_3,
\\
\label{I3dx2}
dx_2&=&\left(\dfrac{\partial x_2}{\partial q_1}\right)_{(2)}dq_1+\left(\dfrac{\partial x_2}{\partial q_2}\right)_{(2)}dq_2+\left(\dfrac{\partial x_2}{\partial x_3}\right)_{(2)}dx_3,
\end{eqnarray}
\end{subequations}
where $q_1=q_1(x_1,x_2,x_3)$ in Eq.~\eqref{I3dq12},
$q_2=q_2(x_1,x_2,x_3)$ in Eq.~\eqref{I3dq2},
$x_1=x_1(q_1,q_2,x_3)$ in Eq.~\eqref{I3dx12},
and $x_2=x_2(q_1,q_2,x_3)$ in Eq.~\eqref{I3dx2},
respectively.
Substituting Eqs.~\eqref{I3dx12} and \eqref{I3dx2} into
Eqs.~\eqref{I3dq12} and \eqref{I3dq2}, we obtain
\begin{eqnarray}
\label{3DJacobiCramer}
dq_1
%&=&
%\left(\dfrac{\partial q_1}{\partial x_1}\right)_{x_2,x_3}\left[\left(\dfrac{\partial x_1}{\partial q_1}\right)_{q_2,x_3}dq_1+\left(\dfrac{\partial x_1}{\partial q_2}\right)_{q_1,x_3}dq_2+\left(\dfrac{\partial x_1}{\partial x_3}\right)_{q_1,q_2}dx_3\right]
%\nonumber\\
%&&+\left(\dfrac{\partial q_1}{\partial x_2}\right)_{x_1,x_2}\left[\left(\dfrac{\partial x_2}{\partial q_1}\right)_{q_2,x_3}dq_1+\left(\dfrac{\partial x_2}{\partial q_2}\right)_{q_1,x_3}dq_2+\left(\dfrac{\partial x_2}{\partial x_3}\right)_{q_1,q_2}dx_3\right]+\left(\dfrac{\partial q_1}{\partial x_3}\right)_{x_1,x_2}dx_3
%\nonumber\\
&=&\left[\frac{\partial q_1}{\partial x_1}\left(\frac{\partial x_1}{\partial q_1}\right)_{(2)}+\frac{\partial q_1}{\partial x_2}\left(\frac{\partial x_2}{\partial q_1}\right)_{(2)}\right]dq_1
\nonumber\\
&+&\left[\frac{\partial q_1}{\partial x_1}\left(\frac{\partial x_1}{\partial q_2}\right)_{(2)}+\frac{\partial q_1}{\partial x_2}\left(\frac{\partial x_2}{\partial q_2}\right)_{(2)}\right]dq_2
\nonumber\\
&+&\left[\frac{\partial q_1}{\partial x_1}\left(\frac{\partial x_1}{\partial x_3}\right)_{(2)}+\frac{\partial q_1}{\partial x_2}\left(\frac{\partial x_2}{\partial x_3}\right)_{(2)}+\frac{\partial q_1}{\partial x_3}\right]dx_3,
\nonumber\\
dq_2
%&=&\left(\dfrac{\partial q_2}{\partial x_1}\right)_{x_2,x_3}\left[\left(\dfrac{\partial x_1}{\partial q_1}\right)_{q_2,x_3}dq_1+\left(\dfrac{\partial x_1}{\partial q_2}\right)_{q_1,x_3}dq_2+\left(\dfrac{\partial x_1}{\partial x_3}\right)_{q_1,q_2}dx_3\right]
%\nonumber\\
%&&+\left(\dfrac{\partial q_2}{\partial x_2}\right)_{x_1,x_3}\left[\left(\dfrac{\partial x_2}{\partial q_1}\right)_{q_2,x_3}dq_1+\left(\dfrac{\partial x_2}{\partial q_2}\right)_{q_1,x_3}dq_2+\left(\dfrac{\partial x_2}{\partial x_3}\right)_{q_1,q_2}dx_3\right]+\left(\dfrac{\partial q_2}{\partial x_3}\right)_{x_1,x_2}dx_3
%\nonumber\\
&=&\left[\frac{\partial q_2}{\partial x_1}\left(\frac{\partial x_1}{\partial q_1}\right)_{(2)}+\frac{\partial q_2}{\partial x_2}\left(\frac{\partial x_2}{\partial q_1}\right)_{(2)}\right]dq_1
\nonumber\\
&+&\left[\frac{\partial q_2}{\partial x_1}\left(\frac{\partial x_1}{\partial q_2}\right)_{(2)}+\frac{\partial q_2}{\partial x_2}\left(\frac{\partial x_2}{\partial q_2}\right)_{(2)}\right]dq_2
\nonumber\\
&+&\left[\frac{\partial q_2}{\partial x_1}\left(\frac{\partial x_1}{\partial x_3}\right)_{(2)}+\frac{\partial q_2}{\partial x_2}\left(\frac{\partial x_2}{\partial x_3}\right)_{(2)}+\frac{\partial q_2}{\partial x_3}\right]dx_3.
\end{eqnarray}
Comparing both sides of Eq.~\eqref{3DJacobiCramer}, we find 
two relevant non-trivial equations:
\begin{eqnarray}
\frac{\partial q_1}{\partial x_1}\left(\frac{\partial x_1}{\partial x_3}\right)_{(2)}+\frac{\partial q_1}{\partial x_2}\left(\frac{\partial x_2}{\partial x_3}\right)_{(2)}&=&-\frac{\partial q_1}{\partial x_3},
\nonumber\\
\frac{\partial q_2}{\partial x_1}\left(\frac{\partial x_1}{\partial x_3}\right)_{(2)}+\frac{\partial q_2}{\partial x_2}\left(\frac{\partial x_2}{\partial x_3}\right)_{(2)}&=&-\frac{\partial q_2}{\partial x_3}.
\end{eqnarray}
By making use of Cramer's rule, we find that
\begin{equation}
\left(\frac{\partial x_1}{\partial x_3}\right)_{(2)}
=
\frac{
\mathscr{Det}
\begin{pmatrix}
\dfrac{\partial q_1}{\partial x_2} & 
\dfrac{\partial q_1}{\partial x_3}
\\
\dfrac{\partial q_2}{\partial x_2} & 
\dfrac{\partial q_2}{\partial x_3}
\end{pmatrix}
}
{
\mathscr{Det}
\begin{pmatrix}
\dfrac{\partial q_1}{\partial x_1} & 
\dfrac{\partial q_1}{\partial x_2}
\\	
\dfrac{\partial q_2}{\partial x_1} & 
\dfrac{\partial q_2}{\partial x_2}
\end{pmatrix}
},
\quad
\left(\frac{\partial x_2}{\partial x_3}\right)_{(2)}
=
-\dfrac{
\mathscr{Det}
\begin{pmatrix}
\dfrac{\partial q_1}{\partial x_1} 
& \dfrac{\partial q_1}{\partial x_3}
\\
\dfrac{\partial q_2}{\partial x_1} & 
\dfrac{\partial q_2}{\partial x_3}
\end{pmatrix}
}
{\mathscr{Det}
\begin{pmatrix}
\dfrac{\partial q_1}{\partial x_1} & 
\dfrac{\partial q_1}{\partial x_2}
\\
\dfrac{\partial q_2}{\partial x_1} & 
\dfrac{\partial q_2}{\partial x_2}
\end{pmatrix}
}.
\label{I3par}%
\end{equation}
%where we have omitted subscripts $x_i,x_j$ ($i\neq j \neq k$)
%for each partial derivative $(\partial q_l/\partial x_k)$ in the right sides of Eq.~\eqref{I3par}.
Then, by making use of Eq.~\eqref{I3par}, we find that 
the factor $\left( \partial q_3 / \partial x_3 \right)_{(2)}$ in the denominator of Eq.~\eqref{I3x3} 

can be expressed as
\begin{eqnarray}
\label{useofchainrule2}
\left(\frac{\partial q_3}{\partial x_3}\right)_{(2)}%&=&
=\frac{\partial q_3}{\partial x_1}\left(\frac{\partial x_1}{\partial x_3}\right)_{(2)}+\frac{\partial q_3}{\partial x_2}\left(\frac{\partial x_2}{\partial x_3}\right)_{(2)}+\frac{\partial q_3}{\partial x_3}
=
\frac{\mathscr{Det}
	\begin{pmatrix}
	\dfrac{\partial q_1}{\partial x_1} & \dfrac{\partial q_1}{\partial x_2} & \dfrac{\partial q_1}{\partial x_3}
	\\
	\dfrac{\partial q_2}{\partial x_1} & \dfrac{\partial q_2}{\partial x_2} & \dfrac{\partial q_2}{\partial x_3}
	\\
	\dfrac{\partial q_3}{\partial x_1} & \dfrac{\partial q_3}{\partial x_2} & \dfrac{\partial q_3}{\partial x_3}
	\end{pmatrix}
}
{\mathscr{Det}
	\begin{pmatrix}
	\dfrac{\partial q_1}{\partial x_1} & \dfrac{\partial q_1}{\partial x_2}
	\\
	\dfrac{\partial q_2}{\partial x_1} & \dfrac{\partial q_2}{\partial x_2}
	\end{pmatrix}
},
\end{eqnarray}
which leads to
\begin{equation}
\label{I3j3j2}
\left|
\left(\dfrac{\partial q_3}{\partial x_3}\right)_{(2)}
\right|
=\frac{\mathscr{G}_3}{\mathscr{G}_2}.
\end{equation}

\section{$\textbf{\textit{n}}$-dimensional case\label{APC}}

The total differentials of $q_1$, $\cdots$, $q_i$, $x_1$, $\cdots$, $x_i$ can be expressed as
\begin{eqnarray}
dq_1&=&\sum_{a=1}^n\frac{\partial q_1}{\partial x_a}dx_a,
\nonumber\\
&\vdots&
\nonumber\\
dq_i&=&\sum_{a=1}^n\frac{\partial q_i}{\partial x_a}dx_a,
\nonumber\\
dx_1&=&\sum_{a=1}^i\left[\left(\frac{\partial x_1}{\partial q_a}\right)_{(i)}dq_a\right]+\sum_{a=i+1}^n\left[\left(\frac{\partial x_1}{\partial x_a}\right)_{(i)}dx_a\right],
\nonumber\\
&\vdots&
\nonumber\\
dx_i&=&\sum_{a=1}^i\left[\left(\frac{\partial x_i}{\partial q_a}\right)_{(i)}dq_a\right]+\sum_{a=i+1}^n\left[\left(\frac{\partial x_i}{\partial x_a}\right)_{(i)}dx_a\right].
\end{eqnarray}
The partial derivative of $q_{i+1}$ with respect to $x_{i+1}$ holding $q_1$, $\cdots$, $q_i$, $x_{i+2}$, $\cdots$, $x_n$ fixed is
\begin{equation}
\label{NDithTotalDerivative}
\left(\dfrac{\partial q_{i+1}}{\partial x_{i+1}}\right)_{(i)}=\sum_{a=1}^i\left[\frac{\partial q_{i+1}}{\partial x_a}\left(\frac{\partial x_a}{\partial _{i+1}}\right)_{(i)}\right]+\frac{\partial q_{i+1}}{\partial x_{i+1}}.
\end{equation}
Substituting $dx_1$, $\cdots$, $dx_i$ into $dq_j$, we obtain
\begin{eqnarray}
\label{NDJacobiCramer1}
dq_j&=&\sum_{a=1}^i\frac{\partial q_j}{\partial x_a}\Bigg[\sum_{b=1}^i\left(\frac{\partial x_a}{\partial q_b}\right)_{(i)}dq_b+\sum_{b=i+1}^n\left(\frac{\partial x_a}{\partial x_b}\right)_{(i)}dx_b\Bigg]+\sum_{a=i+1}^n\frac{\partial q_j}{\partial x_a}dx_a,
\end{eqnarray}
where $1\leq j\leq i$. Because the coefficient of $dx_{i+1}$ in \eqref{NDJacobiCramer1} should be $0$, we obtain the following equation:
\begin{eqnarray}
\begin{pmatrix}
\dfrac{\partial q_1}{\partial x_1} & \cdots & \dfrac{\partial q_1}{\partial x_i}
\\
\vdots & \ddots & \vdots
\\
\dfrac{\partial q_i}{\partial x_1} & \cdots & \dfrac{\partial q_i}{\partial x_i}
\end{pmatrix}
\begin{pmatrix}
\left(\dfrac{\partial x_1}{\partial x_{i+1}}\right)_{(i)}
\\
\vdots
\\
\left(\dfrac{\partial x_i}{\partial x_{i+1}}\right)_{(i)}
\end{pmatrix}
=-
\begin{pmatrix}
\dfrac{\partial q_1}{\partial x_{i+1}}
\\
\vdots
\\
\dfrac{\partial q_i}{\partial x_{i+1}}
\end{pmatrix}.
\end{eqnarray}
By making use of Cramer's rule, we find that
\begin{eqnarray}
\label{NDCramerResult}
\left(\frac{\partial x_j}{\partial x_{i+1}}\right)_{(i)}=-\frac{\mathscr{Det}[(\mathbbm{J}^{-1}_{[i\times i]})^{(j)}(\partial\mathbbm{q}_i)]}{\mathscr{Det}[\mathbbm{J}^{-1}_{[i\times i]}]},
\end{eqnarray}
where
\begin{eqnarray}
\mathbbm{J}^{-1}_{[i\times i]}&=&
\begin{pmatrix}
\dfrac{\partial q_1}{\partial x_1} & \cdots & \dfrac{\partial q_1}{\partial x_i}
\\
\vdots & \ddots & \vdots
\\
\dfrac{\partial q_i}{\partial x_1} & \cdots & \dfrac{\partial q_i}{\partial x_i}
\end{pmatrix},
\nonumber\\
\partial\mathbbm{q}_i&=&
\begin{pmatrix}
\dfrac{\partial q_1}{\partial x_{i+1}}
\\
\vdots
\\
\dfrac{\partial q_i}{\partial x_{i+1}}
\end{pmatrix},
\end{eqnarray}
and $(\mathbbm{J}^{-1}_{[i\times i]})^{(j)}(\partial\mathbbm{q}_i)$ is identical to $\mathbbm{J}^{-1}_{[i\times i]}$ except that the $j$th column is replaced with $\partial\mathbbm{q}_i$.
Substituting \eqref{NDCramerResult} into \eqref{NDithTotalDerivative}, we obtain
\begin{eqnarray}
\label{NDJacobianResult1}
\left(\dfrac{\partial q_{i+1}}{\partial x_{i+1}}\right)_{(i)}&=&\frac{\partial q_{i+1}}{\partial x_{i+1}}-\sum_{j=1}^i\frac{\mathscr{Det}[(\mathbbm{J}^{-1}_{[i\times i]})^{(j)}(\partial\mathbbm{q}_i)]}{\mathscr{Det}[(\mathbbm{J}^{-1}_{[i\times i]})]}\frac{\partial q_{i+1}}{\partial x_j}
\nonumber\\
&=&\frac{1}{\mathscr{Det}[\mathbbm{J}^{-1}_{[i\times i]}]}\sum_{j=1}^{i+1}(-1)^{i+1-j}\mathscr{M}_{i+1j}(\mathbbm{J}^{-1}_{[(i+1)\times(i+1)]})
\nonumber\\
&=&\frac{\mathscr{Det}[\mathbbm{J}^{-1}_{[(i+1)\times(i+1)]}]}{\mathscr{Det}[\mathbbm{J}^{-1}_{[i\times i]}]}.
\end{eqnarray}
Here, the $ij$ minor $\mathscr{M}_{ij}(\mathbbm{A})$ of an $n\times n$ square matrix $\mathbbm{A}$ is the determinant of a matrix whose $i$th row and $j$th column are removed from $\mathbbm{A}$. Hence, Eq. \eqref{NDJacobianResult1} leads to
\begin{equation}
\label{NDJacobianResult2}
\left|\left(\frac{\partial q_{i+1}}{\partial x_{i+1}}\right)_{(i)}\right|=\frac{\mathscr{G}_{i+1}}{\mathscr{G}_{i}}.
\end{equation}

%\end{widetext}

\end{document}